# Ultra-Stretchable Interconnects for High-Density Stretchable Electronics

**Salman Shafqat** [1,†], **Johan P. M. Hoefnagels** [1,†,*], **Angel Savov** [2], **Shivani Joshi** [2], **Ronald Dekker** [2,3] and **Marc G. D. Geers** [1]

[1] Department of Mechanical Engineering, Eindhoven University of Technology, 5600 MB Eindhoven, The Netherlands; s.shafqat@tue.nl (S.S.); m.g.d.geers@tue.nl (M.G.D.G.)
[2] Department of Microelectronics, Delft University of Technology, 2628 CD Delft, The Netherlands; angel.savov@philips.com (A.S.); shivani.joshi@philips.com (S.J.); ronald.dekker@philips.com (R.D.)
[3] Philips Research, High Tech Campus 4, 5656 AE Eindhoven, The Netherlands
* Correspondence: j.p.m.hoefnagels@tue.nl; Tel.: +31-40-2475894
† These authors should be regarded as joint first author.

**Abstract:** The exciting field of stretchable electronics (SE) promises numerous novel applications, particularly in-body and medical diagnostics devices. However, future advanced SE miniature devices will require high-density, extremely stretchable interconnects with micron-scale footprints, which calls for proven standardized (complementary metal-oxide semiconductor (CMOS)-type) process recipes using bulk integrated circuit (IC) microfabrication tools and fine-pitch photolithography patterning. Here, we address this combined challenge of microfabrication with extreme stretchability for high-density SE devices by introducing CMOS-enabled, free-standing, miniaturized interconnect structures that fully exploit their 3D kinematic freedom through an interplay of buckling, torsion, and bending to maximize stretchability. Integration with standard CMOS-type batch processing is assured by utilizing the Flex-to-Rigid (F2R) post-processing technology to make the back-end-of-line interconnect structures free-standing, thus enabling the routine microfabrication of highly-stretchable interconnects. The performance and reproducibility of these free-standing structures is promising: an elastic stretch beyond 2000% and ultimate (plastic) stretch beyond 3000%, with <0.3% resistance change, and >10 million cycles at 1000% stretch with <1% resistance change. This generic technology provides a new route to exciting highly-stretchable miniature devices.

## 1. Introduction

The emerging field of stretchable electronics, especially with its recent advances in high stretchability, has opened a new arena of novel and exciting applications, particularly for medical devices [1–3]. Two main approaches for realizing stretchable electronics (SE) exist: intrinsically stretchable (organic) materials [4–6], and (inorganic) conductor materials made stretchable through inventive mechanisms that convert small strains into a larger global stretch [2,7–9]. In order to realize advanced high-density stretchable electronic miniature devices on a commercial scale, e.g., an inflatable catheter-tip ultrasound camera with variable-zoom functionality for minimally-invasive surgery (Figure 1a), challenges regarding (i) miniaturization; (ii) manufacturability; and (iii) high interconnect stretchability need to be addressed. High-density circuit integration of application-specific integrated circuit (ASIC) devices requires high-density stretchable wiring with width and thickness dimensions in the range of 1 μm, and an overall footprint in the order of tens of micrometers. The required feature sizes are one to three orders of magnitude smaller than the (sub)millimeter-sized serpentines and arches commonly proposed in the literature [2]. This constitutes a real challenge for mesoscopic fabrication procedures such as screen printing and transfer printing, which are frequently used for fabricating large-scale stretchable electronics [2,10–13], while advancements in this area are being made [14]. Conversely, the submicron-sized features are naturally achieved when using the standard fine-pitch photolithography-based IC techniques that are routinely used for the ASIC fabrication. This also warrants the commercial mass production of highly stretchable devices, as long as the process flow utilizes available bulk IC microfabrication equipment and proven CMOS-type process flows. For most (stretchable) electronics applications, having high areal coverage is important, especially for detectors [15]. This calls for a smaller interconnect footprint area and a higher interconnect stretchability. For instance, the inflation of a typical, planar 2D charge-coupled device (CCD) detector array with a fill factor of 90% [10] (i.e., 90% of the surface area covered with rigid, square detector islands) to a hemisphere entails a global strain of 57%, but requires the interconnects between the islands to stretch beyond 1100% (using relation from [16], see Supplementary Materials for details, also illustrated in Figure 1b). For a fill factor of only 50%, still, a stretchability of ~200% is needed. Note that 'stretch' is defined as the global linear strain of the stretchable part of the interconnect (i.e., change in footprint width over initial footprint width (illustrated in Figure 2)). In addition, reliable device operation requires the interconnect structures to accurately recover their original shape upon unloading, which means that the interconnect material should remain below its engineering yield limit. In other words, the key parameter to ensure mechanical reliability is elastic stretchability (defined here as fully recoverable deformation without any visible interconnect shape change upon unloading [17]) and not the ultimate (plastic) stretchability (i.e., stretchability at interconnect fracture in the plastic regime) that is often reported in the literature. Moreover, a large safety factor (on the order of two) is required for cyclic operation below the fatigue limit, as well as for medical applications. Therefore, to unlock the full potential of future high-density SE devices, interconnects that can stretch well beyond 1000% should be combined with a microfabrication solution that warrants the miniaturization and manufacturability of these interconnects and their stretchable devices.

Most of the current stretchability solutions stay (well) below 100% elastic stretch [2]. Many proposed solutions hinge on some sort of serpentine-shaped metal interconnect pattern adhered to a rubber substrate [2], for which the stretching behavior is confined to in-plane deformation. This deformation triggers early plasticity at the interconnect corners, and eventually results in interface delamination, which causes local interconnect rupture [11,12]. In a recent thorough optimization study, the elastic stretchability of substrate-adhered interconnects has been pushed up to 350% [18]. A further huge increase in stretchability calls for a different approach to circumvent such issues as interface delamination and a substrate-induced confinement of deformation modes. Recently, in two milestone studies [13,18], free-standing interconnects have been utilized to boost stretchability. For both approaches, however, integration with standard IC microfabrication has not yet been explored and would be far from trivial, as the processing starts from a polymethyl methacrylate/polyimide (PMMA/PI) layer [13] and a Kapton film [18] as substrate, respectively, while the amount of elastic stretchability is, ~200% and unreported, respectively (as Su et al. discussed in [19]). To achieve advanced high-density SE devices, it would be highly advantageous if the interconnect processing would be fully integrated with (CMOS-type) microfabrication.

Here, the Flex-to-Rigid (F2R) microfabrication platform is adopted and extended to enable CMOS-type processing of highly stretchable (>1000%) interconnects with micron-scale interconnect footprints. F2R is a generic technology that makes CMOS devices flexible by etching the silicon ASICs into separate islands connected by flexible interconnects (formed from the original CMOS-produced interconnects of the ASICs) [20,21]. The strength of the F2R technology has been demonstrated by a flexible (not yet stretchable) ultrasound camera mounted on top of a catheter tip, which was microfabricated in a planar configuration (Figure 1d) and subsequently wrapped around a narrow tip (Figure 1c). Subsequently, this technology was extended using meander-type interconnects, which allow for limited stretchable CMOS-devices [22]. It is important to note that the F2R approach is based on a true post-processing procedure that works on any (pre-processed) CMOS-processed IC wafer, i.e., all 'exotic' processing steps and materials are introduced at the end without affecting the ASIC functionality. By extending the F2R technology to process highly stretchable interconnects, the technological requirements on miniaturization and manufacturability of stretchable interconnects for high-density SE devices are automatically fulfilled. Therefore, the combined challenge consists of (i) an extension of F2R technology to enable free-standing interconnects; (ii) a dedicated stretchable interconnect design that takes into account that the back-end-of-line metallization with a typical thickness of ~0.5 μm is processed into the highly free-standing interconnects; and (iii) a full exploitation of the broad 3D kinematic freedom through an interplay of buckling, torsion, and bending to achieve a reversible stretchability well beyond 1000%.

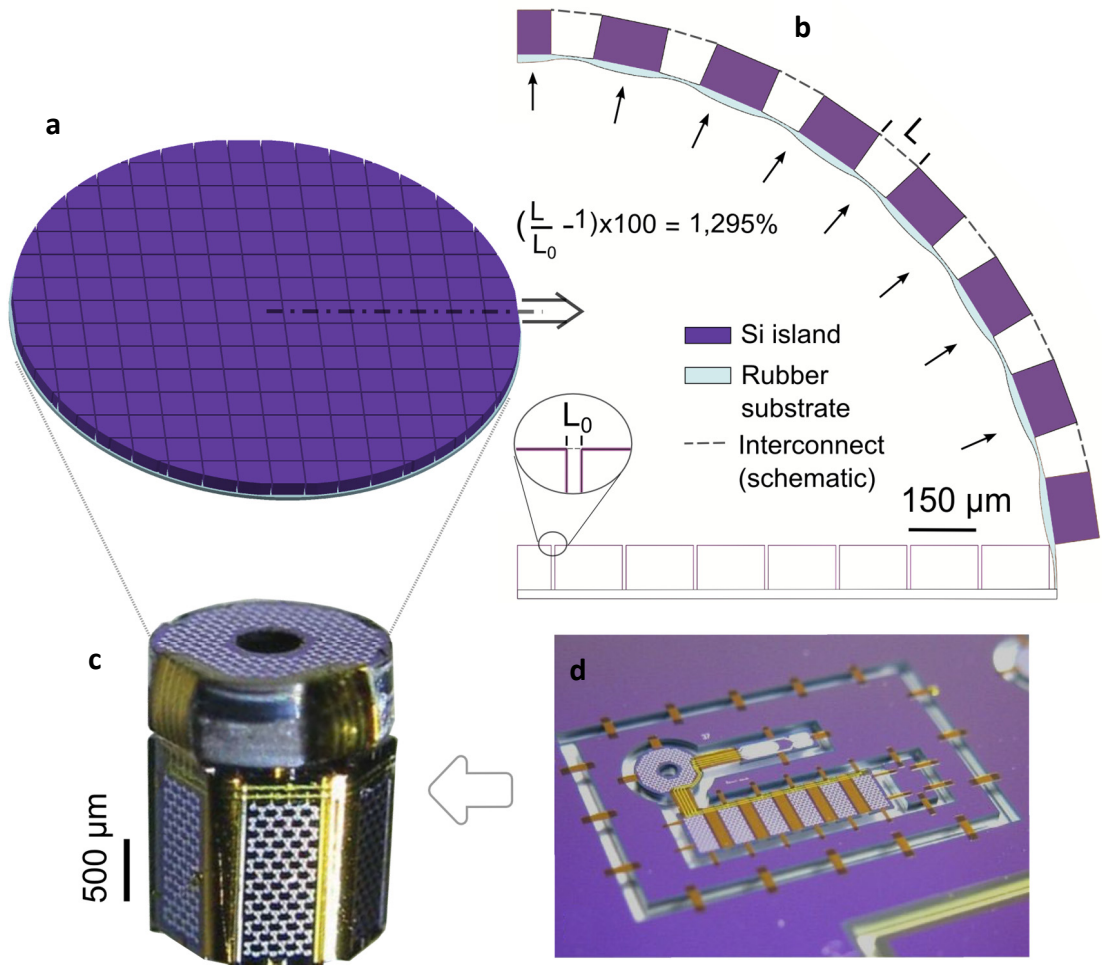

**Figure 1.** Concept illustration of an inflatable catheter-tip ultrasound detector with highly stretchable interconnects (**a**,**b**), versus the current state-of-the-art technology in miniature flexible devices: the capacitive micromachined ultrasound transducer (CMUT) array on a catheter tip produced using the F2R technology (**c**,**d**). © (2013) IEEE. Reprinted, with permission, from [20]. Proposed stretchable detector head, adding a variable-zoom functionality with a large magnification factor by inflating and deflating the stretchable membrane. A flat (deflated) configuration illustrates the rigid silicon islands with a fill factor of 0.9 on a rubber substrate (**a**), and a finite element (FE) simulation of (half of) the cross-section, which demonstrates the requirement of the interconnect to stretch >1200% upon inflation of the detector to a hemisphere with detector islands covering 90% of the membrane area (**b**); (**c**) The fully-assembled CMUT detector with the detector islands bent around the catheter tip; (**d**) The device was microfabricated on a flat standard wafer and suspended by removable polyimide tabs using F2R post-processing technology.

## 2. Interconnect Design and Mechanics

The conceptual design of the free-standing interconnect consists of slender beam elements (see Figure 2 (left)), i.e., the beam thickness is small compared to the beam width and length. This is typical for CMOS-processed (planar) structures. For such beam elements, the maximum elastic tip deflection can be achieved in bending, and more specifically by loading the beam along its thickness direction rather than the (considerably stiffer) width direction. Therefore, the bending beams should rotate in order to align this direction in the global stretch direction, as visualized in Figure 2 (right). To this end, at both island connection points, a slender torsion beam is inserted that rotates the bending beams sideways. Since the structure is initially planar, it will first deform by in-plane opening, which is a deformation mode that quickly exceeds the elastic limit and should be avoided. Therefore, a buckling instability is intentionally triggered, by ensuring that the beam elements are long and thin, to induce a transition from in-plane opening to out-of-plane rotation, well before the onset of plasticity due to in-plane opening. Finally, by connecting the slender bending and torsion beams alternatively at opposite ends, a continuous interconnect structure is formed, as shown in Figure 2 (right). It will be shown that this mechanics-based design of an ultra-compliant, free-standing interconnect activates bending, torsion, and their interplay, and enables it to reach beyond 2000% reversible stretchability.

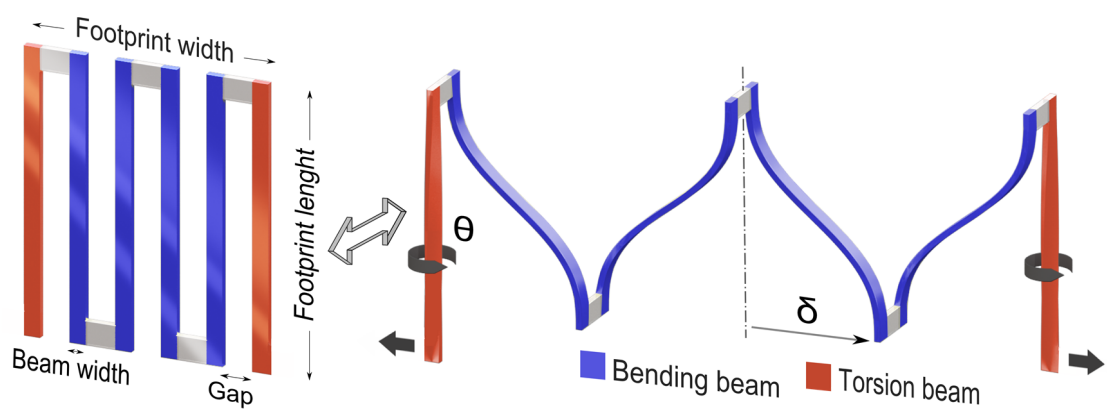

**Figure 2.** Concept illustration of the basic working principle of the free-standing interconnect design. (**Left**) Basic components of the design, with two torsion beams at the corners and multiple bending beams in the middle; (**right**) Basic deformation modes in the stretched state with the corner beams elastically rotated by 90°, which allows the inner beams to elastically bend in the direction of their thickness. $\delta$ and $\theta$ represent the tip deflection of the bending beam and the rotation angle of the torsion beam, respectively.

To ensure that the metal is strained only in the elastic regime, the von Mises yield criterion is used, i.e., the von Mises equivalent stress, $\sigma_{vm,max}$ should remain below the yield stress, $\sigma_y$. The bending beams (with a rectangular cross section) are being deflected as guided cantilever beams. For an individual guided cantilever beam, the equivalent von Mises stress is equal to the maximum normal stress $\sigma_{max}$ along the beam length, which is maximum at the top and bottom surface at the beam's ends (e.g., point b in Figure 3), i.e.,

$$\sigma_{vm,max} = \sigma_{max} \leq \sigma_y \tag{1}$$

The maximum tip deflection $\delta_{max}$ ($\delta$ illustrated in Figure 2) is related to the maximum stress in the beam, i.e., $\sigma_{max}$ using Equation (2) (see Supplementary Materials for details). At the limit where the complete beam is still in the elastic regime ($\sigma_{max} = \sigma_y$), $\delta_{max}$ is proportional to the length squared and inversely proportional to the thickness.

$$\delta_{max} = \frac{(\sigma_{max})l^2}{3Et} = \frac{\sigma_y l^2}{3Et} \tag{2}$$

For an individual torsion beam with a rectangular cross-section, for which plane stress can be assumed in both the thickness and width direction, the maximum equivalent stress equals $\sqrt{3}$ times the maximum shear stress at the beam's surface i.e.,

$$\sigma_{vm,max} = \sqrt{3}\tau_{max} \leq \sigma_y \tag{3}$$

The maximum angle of twist $\theta_{max}$ ($\theta$ illustrated in Figure 2) is related to the maximum shear stress $\tau_{max}$ in the beam using Equation (4) (see Supplementary Materials for details), which shows that at a certain yield stress, $\theta_{max}$ is directly proportional to length $l$ and inversely proportional to thickness $t$.

$$\theta_{max} = \frac{(\tau_{max})l}{ctG} = \frac{\sigma_y l}{\sqrt{3}ctG} \tag{4}$$

where $c$ is a scalar function of the aspect ratio of the beam width $b$ and thickness $t$, which is approximately equal to one for $b/t > 4$, while $b/t > 6.5$ for the current structures due to processing constraints.

The total stretchability of the interconnect, $\epsilon_{global}$ i.e., ($\frac{stretched\ interconnect\ footprint\ width}{initial\ interconnect\ footprint\ width} - 1$) can be approximated as:

$$\epsilon_{global} \approx \frac{(n-2)\delta_{max,c} + 2\delta_{max,t}}{nb + (n-1)g} - 1 \tag{5}$$

where, $n$ is the total number of beams, while $b$ is the beam width and $g$ is the gap between consecutive beams (see Figure 2 (left)). $\delta_{max,c}$ represents the maximum deflection provided by each of the guided cantilever beams, while $\delta_{max,t}$ denotes the deflection of each of the torsion beams in the stretch directions. Unlike the guided cantilever beams, a straightforward analytical equation for $\delta_{max,t}$ is far from trivial, since the torsion beam stiffness is highly non-linear due to its effective area moment of inertia ($I$) varying with increasing angle of twist ($\theta_{max}$). Moreover, the contribution of the torsion beam deflection is only significant for large beam lengths and high global displacements. For example, finite element (FE) simulations show that for Figure 4g, with a relatively high beam length of 100 μm, the contribution of the torsion beam is ~23% of $\epsilon_{global}$ at $\epsilon_{global} = 2040\%$, while at $\epsilon_{global} = 1625\%$, this contribution reduces to ~13% of $\epsilon_{global}$. Furthermore, for typical dimensions, such as those considered here, $\sigma_y$ is reached first in the guided cantilever beams. Thus, it does not dictate the interconnect elastic limit. Therefore, $\delta_{max,t}$ is neglected, resulting in:

$$\epsilon_{global} \approx \frac{(n-2)}{nb + (n-1)g} \frac{\sigma_y l^2}{3Et} - 1 \tag{6}$$

It should be noted that Equation (6) is an approximation, and does not take into account the stress concentrations in the inner corners. Thus, it is not used here to exactly estimate the maximum elastic stretch, for which FE simulations are employed instead. Nonetheless, it provides direct relationships (proportionalities) between maximum stretchability (for a given value of $\sigma_y$) and geometric parameters i.e., beam length $l$, width $b$, thickness, $t$, number of beams $n$, etc., which is invaluable for the design process.

It can be seen from Equations (2), (4), and (6) that in order to achieve higher stretchability, i.e., to postpone the onset of plastic yielding, the thickness of the members (both torsion and bending) should be as small as possible, while the length should be maximized. The width of the beams does not directly influence the maximum stress, and can be chosen to accommodate other design requirements such as interconnect electrical resistance. However, wider members result in a larger initial footprint area, and thus a reduced global stretchability. The gap between adjacent members should be minimal to reduce the initial footprint area, and leave as much space as possible for the ASIC islands. To reduce stress concentrations, the radii of the inside corners of the beam connection points should be at their maximum without increasing the footprint width, i.e., half the gap size. An additional reason to choose the dimensions as small as possible is to benefit from the so-called 'mechanical size effects', which may greatly enhance the yield strength of the aluminum and thus the global elastic stretchability, as is addressed in the following section. On the other hand, a higher length and smaller thickness result in a higher out-of-plane deflection. For example, for a footprint length of 50 μm ($b = 2$ μm and $t = 0.3$ μm), the interconnect deflects out-of-plane by $\pm 7.5$ μm, as seen in Figure 3, while for a length of 100 μm ($b = 2$ μm, $t = 0.3$ μm), the maximum out-of-plane deflection is estimated at $\pm 27$ μm by FE simulation. Therefore, the depth of the trench between the two silicon islands connected by the interconnects should be at least twice that of the maximum out-of-plane deflection, which, for the studied interconnects' geometries is within the range of typically used island heights. Lastly, a key aspect of the design is that the footprint width and footprint length are uncoupled. This results in the freedom to, for example, increase the length of the beams and decrease their thickness in order to increase interconnect stretchability, while also keeping the beam width and gap, and thus the interconnect footprint width fixed, in order to achieve a high fill factor for the ASIC island at the same time. However, the processing-induced curvature, interconnect resistance, and out-of-plane interconnect deflection, which increase with increasing length and decreasing thickness, together or individually will act as the design constrains, on a case-by-case basis.

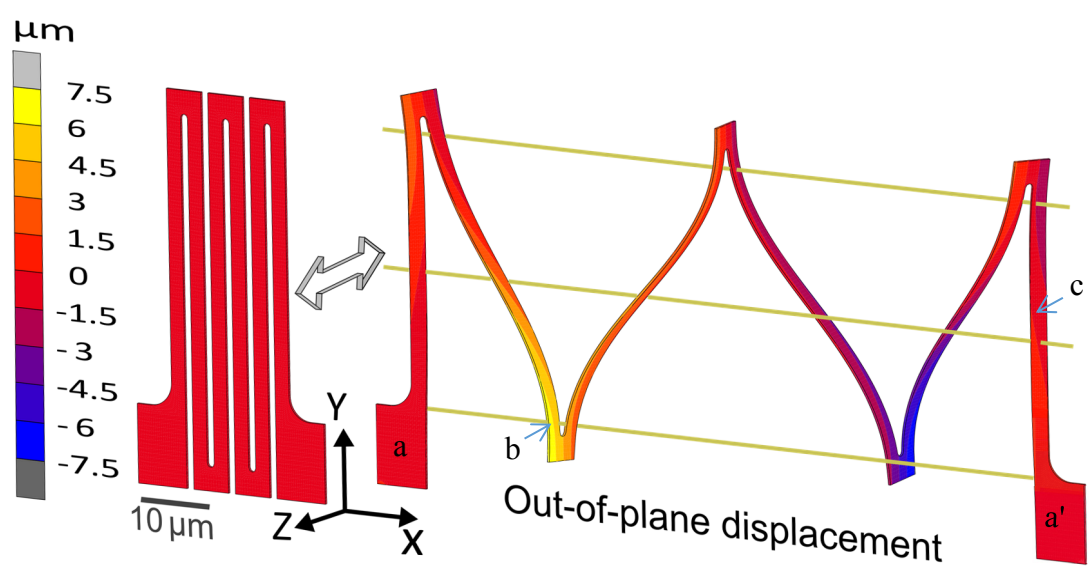

**Figure 3.** Finite element method (FEM) simulation of an interconnect design, showing: (**left**) Initial unloaded structure; (**right**) Stretched interconnect structure, with an out-of-plane displacement field ($U_z$) overlaid on top. Note that the two ends marked a and a' always remain in the initial interconnect plane (i.e., $U_z = 0$), along with the three parallel gold-colored solid lines. Point b and c denote the location of $\sigma_{max}$ and $\tau_{max}$, respectively.

## 3. Results

Aluminum test structures for accurate micromechanical testing (see Appendix B for details) inside a scanning electron microscope were fabricated using the F2R processing scheme (see Appendix A for details). Figure 4a shows a 100 μm-long test structure with four inner members in the initial (load-free) state, which clearly shows an out-of-plane curvature due to processing-induced residual stresses, as is common in microfabrication. As predicted, at ~10% global stretch, the initial regime of in-plane deformation started to transition to out-of-plane rotation of the structure, while ~45° out-of-plane rotation occurs at 190% global stretch, see Figure 4c. Note that due to the (processing-induced) out-of-plane curvature, along with the small width and thickness of the beams, the alignment of the thickness direction of the bending beams with the loading direction after rotation only becomes clear upon close inspection.

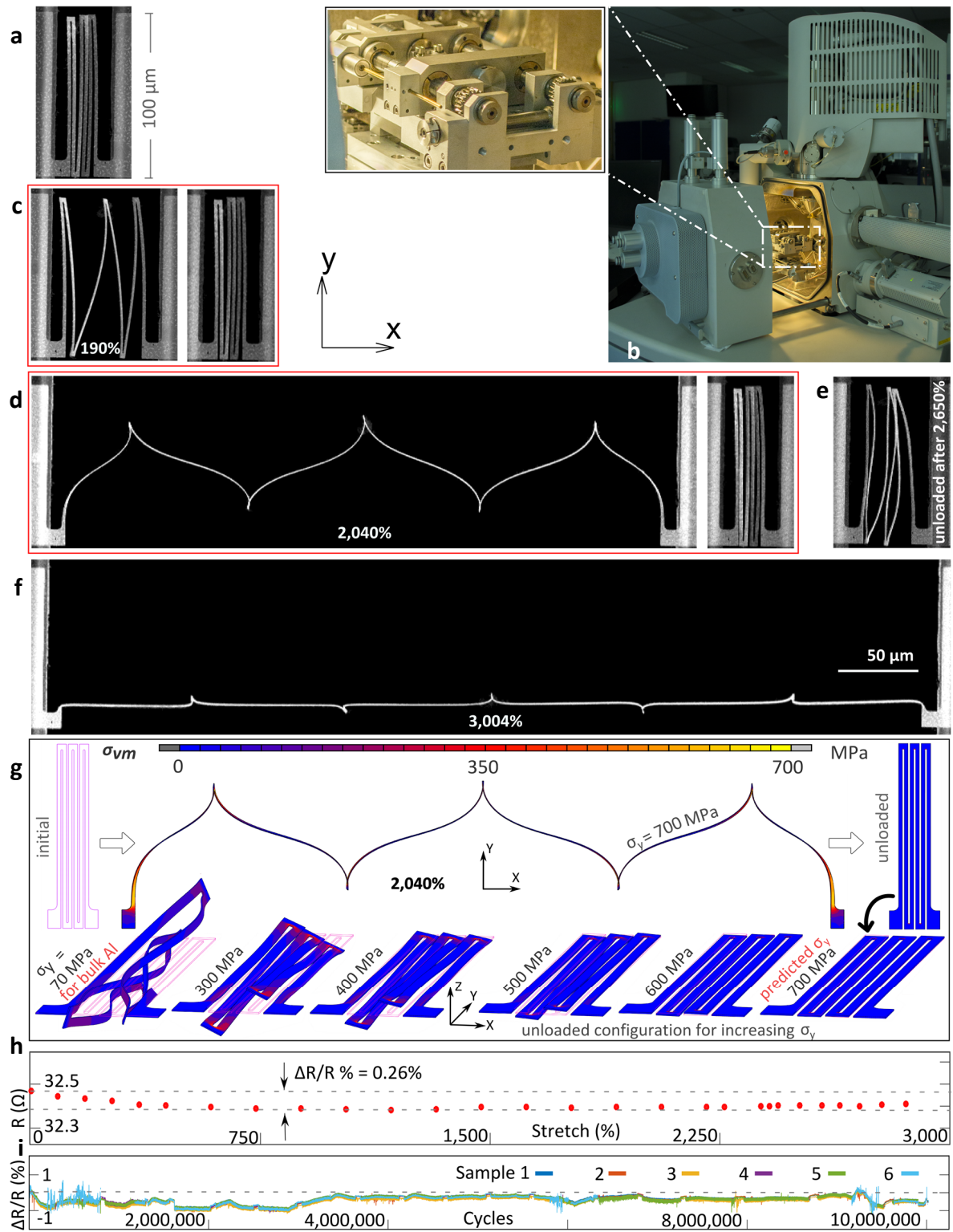

**Figure 4.** Selected experimental and numerical results. (**a**) The initial (load-free) state of the interconnect structure with a beam length, width, and thickness of 100, 2, and 0.3 µm, and gap of 1 µm; (**b**) Microtensile stage placed inside a large chamber scanning electron microscope (SEM), with a magnified view of the tensile stage; (**c**,**d**) Demonstration of reversible stretchability: in each stretching cycle, the interconnect structure is stretched to an incrementally higher global strain (left figure), while the shape after unloading (corresponding right figure) should be compared with the initial configuration; (**e**) Beyond 2040% global stretchability, plastic deformation sets in, which becomes clearly visible after unloading from the 2650% stretched state; (**f**) The ultimate (plastic) stretchability is reached beyond 3000% global strain. Exactly the same deformation behavior was observed for the other four parallel structures in the field of view (Figure S1 in the Supplementary Materials section);

(**g**) finite element method (FEM) elastoplastic simulation at 2040% global strain (compare with (**d**)), for a yield stress of 700 MPa: equivalent von Mises stress overlaid on the stretched configuration (left) and subsequent unloaded configuration (right); (below) Unloaded configurations for increasing yield strength values, showing that the yield stress is at least $\sigma_y$ = (700 ± 100) MPa; (**h**,**i**) Electrical resistance measurements (**h**) for a quasi-static stretch to 3000% and (**i**) during 10 million cycles of 1000% elastic stretch.

To study the elastic stretchability, at each loading step, with increasing applied displacement, the sample is brought back to its unloaded configuration, after which possible permanent shape changes are analyzed in detail. When the structure is unloaded after 190% stretch, no systematic shape change was observed, which demonstrates that the stretchability is fully reversible. After increased loading, the corner members and thus also the middle bending members rotate further out of plane. At ~1500% stretch, the middle members have completely rotated by ~90°, and further deformation is accommodated completely by bending in the stretch direction, as indented by the design. Note that even for the case of Figure 4d, no permanent shape change could be detected, showing that the structure can be elastically stretched up to 2040%.

Elastic reversibility is lost at 2265% global stretch, where for the first time a minute shape change could be observed in the corners of the structure, where the highest stress occurs. The shape change only becomes clearly visible after unloading from the 2650% stretched state, as seen in Figure 4e. Upon further deformation, the structure stretches into an almost straight wire at 3004% of global stretch, see Figure 4f. All of the structures remained intact, while the four parallel structures within the field of view (Figure S1 in the Supplementary Materials) demonstrated exactly the same deformation behavior (including the minor asymmetry due to the processing-induced curvature). This demonstrates the robustness of the interconnect structures due to their ultra-flexible nature. In fact, contrary to their fragile appearance, the interconnects prove to be quite immune to rough handling of the wafer and test chip.

Elastoplastic FEM simulations were performed (see Appendix C for detials) for the 100 μm-long interconnect structure. Figure 4g shows the deformed shape at 2040% global stretch, and the corresponding shape after unloading. Even though the processing-induced curvature is not included in the simulations, the deformed shape shows excellent agreement with the experiments (Figure 4d), which supports our conclusion that the elastic stretchability indeed exceeds 2040%. Also, the local curvatures due to the bending of the inner members and the torsion and bending of the outer members are predicted accurately, which provides further confirmation that the underlying mechanics principles exploited in the design are valid for these microfabricated free-standing structures.

The electrical resistance measurement (Figure 4h) shows that (i) the absolute value (32.5 Ω) varies by 10% from the theoretical value based on the interconnect geometry, and (ii) the resistance remains constant within 0.26% up to an extreme 'plastic' stretch of 2860%, which proves that the electrical behavior of the interconnects is very stable. More importantly, the interconnects exhibit excellent fatigue response. As seen in Figure 4h, all six parallel interconnects survived 10 million cycles at a maximum cyclic stretch of 1000% without failure (see video in Supplementary Materials section), with the resistance remaining constant within 1% and no sign of plasticity. This confirms the goal of achieving a highly reproducible, high elastic stretchability that could ensure reliable device operation.

The onset of plasticity, which manifests as a permanent shape change in an interconnect (see Figure S2 in the Supplementary Materials for details) is predicted by the yield strength of the material. However, at submicrometer scales, the yield strength can significantly vary (and typically increase) from bulk material characteristics due to well-known 'mechanical size effects', which are commonly referred to as 'smaller is stronger' [23–25]. The yield strength of the interconnect material is estimated by qualitatively fitting the deformed shape of the FEM simulation onto the experiment shape by perturbing the yield strength until the simulated unloaded configuration at the elastic limit (i.e., 2040%, see (Figure 4h)) is similar to the experimental unloaded configuration (this procedure is similar to inverse characterization methods reported in the literature [26–29] for MEMS and thin films). An unloaded configuration after a stretch of 2040% without any discernable shape change is achieved only when a yield strength of 700 (±100) MPa is used, which is 10 times higher than that of bulk unalloyed aluminum (~70 MPa), as seen in Figure 4h. Since the maximum stress is found in a small volume of ~0.1 μm³ (near the surface at the bottom of the torsion beams and near the surface in the inner corners), the high yield strength likely results from the aforementioned 'mechanical size effects'. The statistical size effect in polycrystalline aluminum samples with similar dimensions, even in pure tension, has been reported in the literature [30]. Furthermore, the three most prominent strengthening effects in metals are due to strain gradients [24,31], dislocation starvation [24,32], and constrained boundary layers [24]. Our miniature interconnect structures have a native oxide layer, and are extremely bent with highly localized stresses and high strain gradients in the inner corners and submicron-sized grains. Therefore, the individual role of each of these effects cannot be trivially assessed. A detailed study into these size effects is of high scientific interest and may be utilized to further enhance the elastic stretchability. It should be noted that the goal here was not to precisely determine the yield strength of the interconnect material, but rather to show that the yield strength can vary substantially from bulk properties (due to size effects), which can be exploited to boost interconnect stretchability.

## 4. Conclusions and Discussion

To conclude, in this work, a new route is opened towards standard CMOS processing of ultra-stretchable, free-standing interconnects with a micron-sized footprint, which could enable extreme device stretchability. This could be realized by reserving areas on a CMOS-processed wafer upon which the free-standing interconnects are fabricated to connect the individual devices, followed by through wafer etching to separate the devices and release the interconnects, similar to proposals in [20,22]. Such a process scheme keeps the thermal budget below 400 °C, which is a key requirement if CMOS devices are being post-processed. Furthermore, the starting point of the process is a Si-substrate without any specific mechanical or electrical requirements for the substrate. Hence, any CMOS wafer, e.g., a commercially available ASIC wafer, can serve as a substrate.

Further steps towards actual applications are to be taken next. High-density SE devices typically require multi-level wiring between the ASIC islands. The adopted F2R processing scheme is already capable of producing multi-level wiring, which does require the interconnects to be electrically isolated. This could be easily achieved by adding a single processing step to the F2R processing scheme, i.e., deposition of a very thin conformal coating (e.g., of parylene). Moreover, for a 2D (initially flat) detector array conforming to a curvilinear surface (Figure 1b), the ASIC islands connected by the interconnect show some degree of relative displacement perpendicular to the main interconnect stretch direction, as well as relative rotations (currently being studied), which need to be accommodated by the interconnect.

The discussed 100 μm-long interconnects yield an exceptional elastic stretchability of >2000%, which enables breakthrough applications such as the variable-zoom catheter-tip ultrasound detector (Figure 1). Moreover, the ultimate (plastic) stretch of >3000% is also highly interesting for promising one-time stretchable devices [33]. For instance, a CCD-detector stretched into an almost full sphere that mimics a fly's eye and can 'see' in all directions (such as devices reported in [34,35]), can serve as a lightweight omni-directional camera on top of bug-like miniature drones.

**Supplementary Materials:** The following are available online at www.mdpi.com/2072-666X/8/9/277/s1, Figure S1: Parallel interconnects, Figure S2: Equivalent accumulated plastic strain and Video S1: Fatigue test.

**Acknowledgments:** This work was supported by the Vidi funding of J.P.M.H. (project number 12966) within the Netherlands Organization for Scientific Research (NWO). This work was also partially funded by project T62.3.13483 in the framework of the Research Program of the Materials innovation institute (M2i) (www.m2i.nl), and in the framework of the ECSEL JU Project InForMed, grant number 2014-2662155 (www.informed-project.eu).

**Appendix A**

**Fabrication**: The fabrication process starts with the growth of 1 μm and 5 μm PECVD $SiO_2$ on the front and backside respectively of a 150-mm double side polished 400 μm thick Si substrate. First, the oxide layer on the back side is patterned by photolithography and dry etching, forming the hard etch mask for the through-wafer etching required at a later stage. Next, a 300 nm thick layer of aluminum is sputter deposited on the front side of the wafer. A photoresist is spin coated (90 °C soft baked), exposed and developed. Using this resist mask, the aluminum is dry etched defining the desired interconnect patterns along with the re-routing layer and the bond pads. After etching, the resist mask is removed in $O_2$ plasma. A 5.2 μm thick layer of polyimide (PI 2611) is spin coated (soft baked on a hotplate 120 °C for 5 min) on top of the Al structures and cured in nitrogen atmosphere at 275 °C for 2 h. A layer of Al is sputtered (200 nm) on top of the cured polyimide. A layer of photoresist (HPR504) is spin coated, exposed by stepper lithography and developed. The top Al is wet etched with the resist serving as a mask. The patterned Al layer will serve as a hard etch mask for the patterning of the polyimide at the end of the process. The interconnects are next released from the Si substrate by back side Deep Reactive Ion Etching (DRIE) for 25 min with the aid of the 5 μm hard etch PECVD Oxide mask. The final release step is done when the PI is etched selectively from the interconnects from the top using the patterned Al hard etch mask.

**Appendix B**

**Experiments**: Proof of principle interconnect test structures suspended between two silicon islands, resembling actual applications, were fabricated using the F2R processing scheme. The free-standing interconnects were fabricated on a (16 mm × 5 mm) test chip, as shown in Figure A1. To maximize space, six parallel interconnect structures are processed on each test chip, which can be simultaneously mechanically tested. These test structure were tested with a commercial (Kammrath & Weiss GmbH, Dortmund, Germany) microtensile stage designed for in-situ scanning electron microscope (SEM) (Figure 4b). The test chip was glued onto two clamps with a flat acrylic top surface using UV-curable glue by Loxeal®. The in-situ experiments were performed inside a FEI Quanta 600 FEG-SEM under high vacuum, in secondary electron imaging mode. The four-probe electrical resistance measurements were performed using integrated electrical probes that make contact with bonding pads connected to the interconnects on the test chip, while the deformation state of the interconnects was visualized with an optical microscope. The fatigue measurements were also performed under the optical microscope with cyclic loading applied at 10 Hz. After every 300 cycles, the electrical resistance of the interconnects was measured to characterize interconnect failure. Optical images of the interconnects were obtained every million cycles.

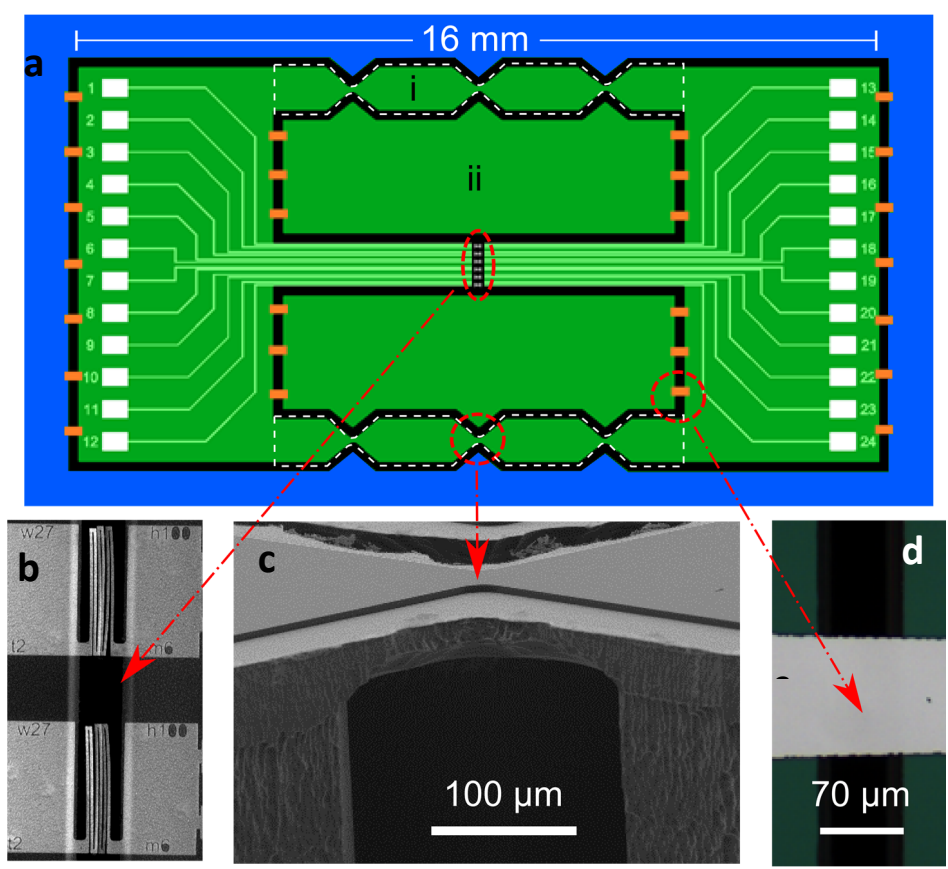

**Figure A1.** Design and details of the fabricated test chip with highly stretchable interconnects. (**a**) The design of the (16 mm × 5 mm) chip (in green), which is suspended in the silicon wafer (in blue) by polyimide tabs (in orange). Notched silicon columns (**i**) are processed to provide rigidity to the structure during processing and handling. Sacrificial silicon islands (**ii**) are created and suspended by similar polyimide tabs to reduce etching time; (**b**) Magnified view of two parallel free-standing interconnects; in total there are six parallel interconnect structures, processed in each test chip; (**c**) Magnified side-view of a (200 μm wide, 100 μm thick, double V-notched) silicon bridge inside the bottom 'column' of the test chip; (**d**) Magnified top-view of a polyimide tab, which can easily be locally melted to singulate the test chip.

## Appendix C

**Simulations**: The elastoplastic FE simulations were performed using a commercial FEA package, Marc Mentat 2014®. 20-node quadratic brick elements were used to model the geometry, with three elements through the thickness to accurately capture the bending behavior. A non-linear solution scheme was used to capture the large displacements upon stretching as well as the buckling bifurcation in the initial phase of loading. To assist the structure to buckle, a minute perturbation force of $10^{-10}$ N was applied to a corner node in the middle member of the structure and removed after the bifurcation point has been passed. Bulk elastic properties of pure aluminum were used for the material properties, i.e., a Young's modulus of E = 69 GPa and a Poisson's ratio of $\nu$ = 0.33. Moreover, a standard Swift power law relationship was used to describe the hardening behavior with a strength coefficient (K) varying from 0.278 GPa ($\sigma_y$ = 70 MPa) to 10.1 GPa ($\sigma_y$ = 700 MPa) and strain-hardening exponent of $n$ = 0.2, as common for aluminum.